\documentclass[10pt,twocolumn,aps,prl,showpacs,groupedaddress]{revtex4}
\usepackage{epsfig}
\usepackage{float}
\usepackage{color}

\begin{document}

\title{Remarkably robust and correlated coherence and antiferromagnetism in (Ce$_{1-x}$La$_x$)Cu$_2$Ge$_2$}

\author{H. Hodovanets$^{1,2}$, S. L. Bud'ko$^{1,2}$, W. E. Straszheim$^{1}$, V. Taufour$^{1,2}$, E. D. Mun$^{2}$, H. Kim$^{2}$, R. Flint$^{1,2}$, and P. C. Canfield$^{1,2}$}

\affiliation{$^1$Ames Laboratory, Iowa State University, Ames, Iowa 50011, USA}
\affiliation{$^2$Department of Physics and Astronomy, Iowa State University, Ames, Iowa 50011, USA}

\begin{abstract}
We present magnetic susceptibility, resistivity, specific heat, and thermoelectric power measurements on (Ce$_{1-x}$La$_x$)Cu$_2$Ge$_2$ single crystals (0 $\leq x\leq$ 1). With La-substitution, the antiferromagnetic temperature $T_N$ is suppressed in an almost linear fashion and moves below 0.36 K, the base temperature of our measurements for $x>$ 0.8. Surprisingly, in addition to robust antiferromagnetism, the system also shows low temperature coherent scattering below $T_{coh}$ up to $\sim$ 0.9 of La, indicating a small percolation limit $\sim$ 9$\%$ of Ce. $T_{coh}$ as a function of magnetic field was found to have different behavior for $x<$ 0.9 and $x>0.9$. Remarkably, $(T_{coh})^2$ at $H$ = 0 was found to be linearly proportional to $T_N$. The jump in the magnetic specific heat $\delta C_{m}$ at $T_N$ as a function of $T_K/T_N$ for (Ce$_{1-x}$La$_x$)Cu$_2$Ge$_2$ follows the theoretical prediction based on the molecular field calculation for the $S$ = 1/2 resonant level model.
\end{abstract}


\maketitle

Dilution studies of the Kondo lattice provide a unique probe to understand the interrelation between Kondo coherence and magnetic order. In a dilution study of the antiferromagnetically (AFM) ordered Kondo lattice (Ce$_{1-x}$La$_x$)Cu$_2$Ge$_2$, we find a remarkably wide region of antiferromagnetic order and Kondo coherence up to $x$ = 0.8 and $x$ = 0.9, respectively, along with an unexpected scaling of $T_N \sim (T_{coh})^2$. This wide region appears to contradict current theoretical predictions for Kondo coherence alone, which state that coherence vanishes for much smaller $x$ \cite{Tesanovic1986}, giving rise to either a broad region of non-Fermi liquid \cite{Kaul2007} or a Lifshitz transition \cite{Lifshitz1960,Burdin2013}. Our findings suggest that, in this system, magnetic correlations actually reinforce the Kondo coherence.

As a result of competition between the Kondo effect and Ruderman-Kittel-Kasuya-Yosida (RKKY) interaction, Kondo lattices display a variety of ground states (long-range magnetic order, unconventional superconductivity, non-Fermi liquid, \textit{etc}. \cite{Stewart1984,Stewart2001,Stewart2006,Stockert2011}) and are characterized by multiple energy scales (antiferromagnetic $T_N$ or superconducting $T_c$ ordering temperature, the single-ion Kondo temperature $T_K$, the coherence temperature $T_{coh}$, and the crystal electric field (CEF) splitting). 
Ce-based compounds, both in coherent and diluted regimes, have been studied for more than four decades with the hope of understanding how coherence develops with an increas of the Kondo impurity concentration (Refs. \cite{Samwer1976,Bredl1978,Felsch1978,Lin1987,Sampathkumaran1989,Garde1994,Hewson1997,Nakatsuji2002,Pikul2012} and references therein). For example, (Ce$_{1-x}$La$_x$)Pb$_3$ shows coherence up to $x$ = 0.15 and single-ion Kondo scaling for a surprisingly wide range of $x$ and $T$ ($T_K$ is the same for these concentrations) \cite{Lin1987,Schlottmann1989}. In the study of (Ce$_{1-x}$La$_x$)Ni$_2$Ge$_2$, the coherence was found up to $x$ = 0.4 with impressive single-ion Kondo scaling in the coherent Fermi-liquid as well as diluted regimes \cite{Pikul2012}. 
 
Based on analysis of La dilution of CeCoIn$_5$ (for which $T_K$, $T_{coh}$, and CEF are well separated), a two-fluid description of the Kondo lattice was put forward \cite{Nakatsuji2004}. It proposes two different energy scales for the Kondo lattice: characteristic temperature $T^*$ ($T^*$=$T_{coh}$ for non-diluted, parent compound) that governs the intersite coupling of the $f$ shells in the coherent Kondo lattice and the concentration-independent single-ion $T_K$, responsible for the on-site 4$f$- conduction-electron hybridization. $T_{coh}$ for this system was observed up to $x\sim$ 0.4 \cite{Nakatsuji2002}.

In this work, we study La dilution of the Kondo lattice compound CeCu$_2$Ge$_2$ where $T_N\sim$ 4 K \cite{Rauchschwalbe1985,Felten1987,Boer1987} and the two excited CEF levels at $\Delta E_1\sim$ 197 K and $\Delta E_2\sim$ 212 K \cite{Loewenhaupt2012} are well separated from the ground state doublet. 
%----------------------------------------------------------------------------------------------------------
%----------------------------------------------------------------------------------------------------------
The measurements were performed on single crystals grown by the high temperature flux method \cite{Canfield1992,Canfield2010,Canfield2001}. The actual concentrations of La or Ce were assessed by wavelength dispersive x-ray spectroscopy (WDS) and the results of the Curie-Weiss fits of the temperature dependent susceptibility. The WDS values of La/Ce concentrations will be used throughout the text if not specified otherwise. La concentrations will be denoted by $x$ and Ce concentrations will be denoted by $y$=1$-x$ to avoid confusion. The details of samples growth, evaluation of La concentrations and measurement techniques can be found in the Supplemental Material \cite{SM}.

%----------------------------------------------------------------------------------------------------------
%----------------------------------------------------------------------------------------------------------

\begin{figure}[!t]
\centering
\includegraphics[width=1\linewidth]{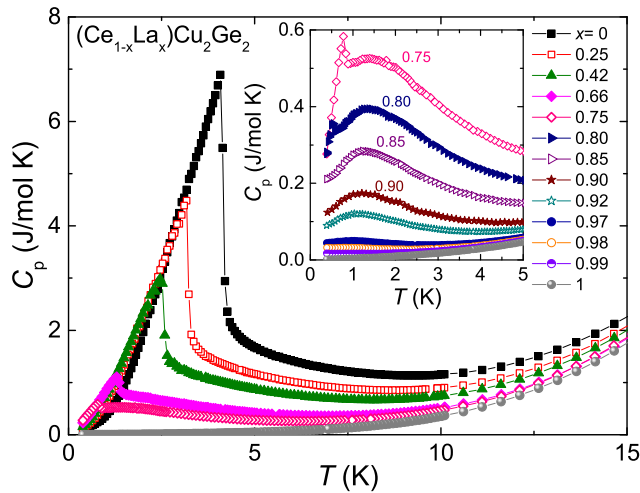}
\caption{\footnotesize (Color online) Specific heat $C_p(T)$ data of (Ce$_{1-x}$La$_x$)Cu$_2$Ge$_2$ single crystals. The inset shows enlarged low-temperature data for 0.75 $\leq x\leq$ 1. The data for $x$ = 0.75 are shown in both graphs for clarity.}
\label{Cp-122}
\end{figure}

An almost classic, mean-field-like second order AFM transition is clearly seen in the specific heat $C_p(T)$ data for CeCu$_2$Ge$_2$ (Fig. \ref{Cp-122}). As the amount of La is increased, the AFM transition moves to lower temperature and is still clearly observable for $x$ = 0.80. When $T_N$ is suppressed enough, in addition to the AFM ordering, a broad maximum appears in the specific heat data starting from $x$ = 0.75 (inset to Fig. \ref{Cp-122}), the position of which shifts slightly to lower temperatures as the La concentration is further increased. The maximum becomes almost indiscernible for $x$ = 0.99. This maximum is associated with the Kondo temperature $T_K$ of a single-ion Kondo impurity ($T_K>T_N$) \cite{SM}.

\begin{figure}[t]
\centering
\includegraphics[width=1\linewidth]{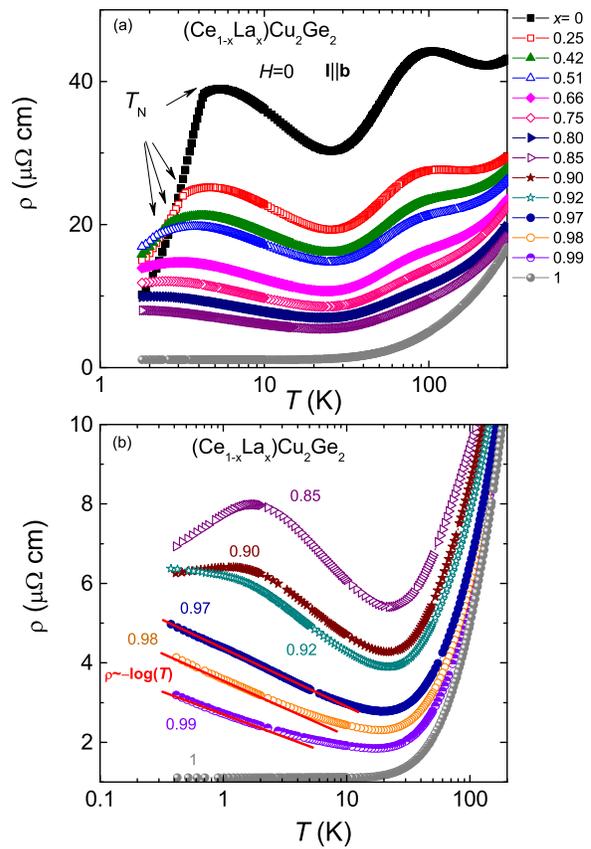}
\caption{\footnotesize (Color online) (a) and (b) The zero-field, in-plane (I$\|$\textbf{b}), temperature-dependent resistivity $\rho(T)$ data of (Ce$_{1-x}$La$_x$)Cu$_2$Ge$_2$ single crystals on a semi-logarithmic plot. The data for $x$ = 0.85 is shown in both panels for continuity.}
\label{all-log}
\end{figure}

A hallmark of the single-ion Kondo effect is the minimum and lower-temperature logarithmic dependence of the resistivity data. The zero-field, temperature dependent, in-plane, resistivity $\rho(T)$ data of (Ce$_{1-x}$La$_x$)Cu$_2$Ge$_2$ are shown on a semi-logarithmic plot in Fig. \ref{all-log}. For CeCu$_2$Ge$_2$, the $\rho(T)$ data exhibit a broad
maximum at $\sim$ 100 K associated with a thermal depopulation of the exited CEF levels as the temperature is decreased. At lower temperatures, the $\rho(T)$ plot shows a second broad maximum corresponding to a crossover from incoherent to coherent scattering of the electrons on the magnetic moments at $T_{coh}\sim$ 5.5 K, characteristic of that of Kondo lattice compounds. The maximum is followed by (and actually truncated by) a kink corresponding to the AFM transition. As the amount of La is increased, the AFM transition moves to lower temperatures. The kink, corresponding to the AFM transition, becomes less discernible. Most intriguingly, the truncated maximum, at $T_{coh}$ for CeCu$_2$Ge$_2$, evolves into a broad maximum and remains present up to $x$ = 0.90, Fig. \ref{all-log}(b). For $x$ = 0.92, the resistivity data tend to saturation at the lowest temperature measured. This behavior in the resistivity is reminiscent of the single-ion Kondo impurity. For the three smallest Ce concentrations, the resistivity data display the minimum followed by a $-$log($T$) dependence upon cooling to the lowest temperature. It is worth pointing out that the slightly temperature dependent minimum at $\sim$ 20 K in the resistivity data is observed for all samples containing Ce. $T_{min}$ is proportional to the concentration of Ce, $y^{1/5}$, only for 0.01 $\leq y\leq$ 0.08 (see the Supplemental Material \cite{SM}) consistent with the single-ion Kondo impurity effect.

\begin{figure}[tbh]
\centering
\includegraphics[width=1\linewidth]{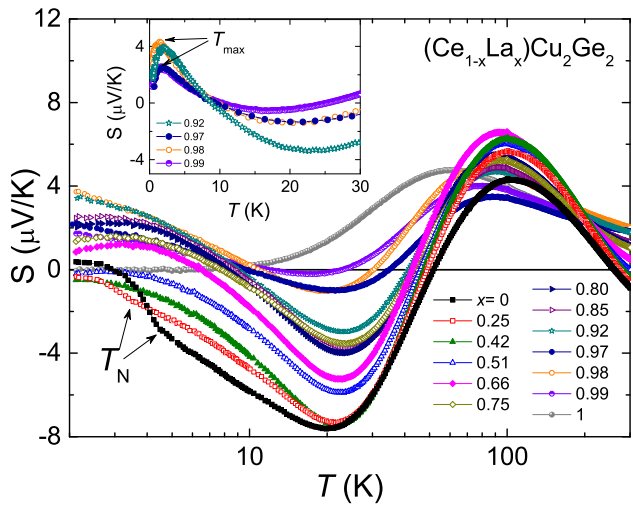}
\caption{\footnotesize (Color online) The zero-field, temperature-dependent thermoelectric power $S(T)$ of (Ce$_{1-x}$La$_x$)Cu$_2$Ge$_2$ single crystals. $S(T)$ of (Ce$_{1-x}$La$_x$)Cu$_2$Ge$_2$ (0.92 $\leq x\leq$ 1) single crystals at lower temperatures is shown in the inset. $\nabla T\|$\textbf{b}.}
\label{allTEP}
\end{figure}

Thermoelectric power (TEP) can also provide information about the $T_{coh}$ and $T_K$ characteristic energy scales. Temperature-dependent thermoelectric power $S(T)$ data for (Ce$_{1-x}$La$_x$)Cu$_2$Ge$_2$ single crystals are shown in Fig. \ref{allTEP}. The broad peak observed for LaCu$_2$Ge$_2$ at $\sim$ 75 K ($\sim$ 0.2$\times\Theta_D$) is probably due to the phonon drag contribution expected at 0.1$-$0.3$\Theta_D$ \cite{Blatt1976,SM}. For all samples containing Ce, a broad, high-$T$ maximum due to (i) the thermal depopulation of the two excited CEF doublets \cite{Loewenhaupt2012} as the temperature is lowered and (ii) possibly phonon drag contribution is observed around 100 K. Since the energy separation between those two excited CEF levels is small, only one maximum at high temperatures is seen in the TEP measurements. The position of this maximum is almost unaffected by La substitution.

The TEP data of LaCu$_2$Ge$_2$ are positive over the whole temperature range measured. However, 0.01 of Ce is enough to change the functional dependence of the TEP below $\sim$ 24 K: the TEP for $x$ = 0.99 crosses zero twice by going through a low-T minimum and has a low-T maximum at $\sim$ 0.6 K (see inset to Fig. \ref{allTEP}). Such TEP behavior is expected for the Ce single-ion Kondo impurity \cite{Zlatic2003,Zlatic2005,Zlatic2007}. For the highly La diluted samples, this low-temperature maximum is believed to correspond to the single-ion impurity Kondo temperature $T_K$. As the amount of Ce is increased, the absolute value of $S_{min}$ increases as well probably reflecting the amount of Ce ions and increased scattering associated with the increase of Ce.

\begin{figure}[t]
\centering
\includegraphics[width=1\linewidth]{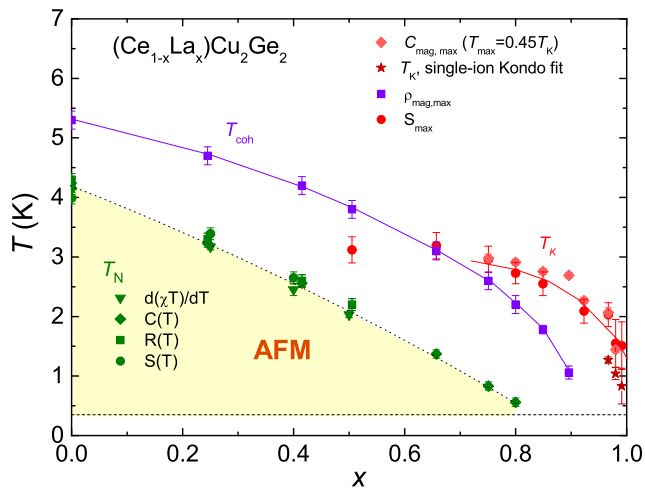}
\caption{\footnotesize (Color online) $T-x$ phase diagram for (Ce$_{1-x}$La$_x$)Cu$_2$Ge$_2$ single crystals. Lines are guides to the eye. The horizontal line at 0.36 K is the lowest base temperature of the measurements. The data for magnetization measurements and $T_K$-values, estimated using Schotte and Schotte model of the specific heat data, can be found in the Supplemental Material \cite{SM}. Kondo temperature $T_K$ was also estimated from the specific heat data by using $T_{max}$ = 0.45 $T_K$ criterion \cite{Desgranges1982} (see text for details).}
\label{PDCe}
\end{figure}

The $T-x$ phase diagram for (Ce$_{1-x}$La$_x$)Cu$_2$Ge$_2$, Fig. \ref{PDCe}, shows the characteristic temperatures and energy scales as a function of La concentration. $T_N$ (as determined from specific heat, magnetization, resistivity and TEP measurements) decreases almost linearly with $x$ and moves below the base temperature of 0.36 K or disappears for $x>$ 0.80 and $T_{coh}$ extends down to $x$ = 0.90. The low-temperature maxima in the TEP data seem to coincide with the $T_K$-values estimated from the specific heat data using a $T_{max}$ = 0.45 $T_K$ criterion \cite{Desgranges1982} rather well, here $T_{max}$ is the temperature where the maximum occurs. The $T_K$ values estimated using the Schotte and Schotte single-ion Kondo model fit \cite{SM,Schotte1975} of the specific heat data although lower, are still within the error bars of the ones estimated using the $T_{max}$ = 0.45 $T_K$, criterion. The decrease of the Kondo temperature from $\sim$ 4 K ($x$ = 0) to $\sim$ 1 K ($x$ = 0.99) upon La substitution is consistent with the unit cell volume increase with $x$ in terms of the Doniach phase diagram \cite{Doniach1977,Coleman2007} \textit{i.e.} the system is tuned away from a quantum critical point (QCP). However, La-substitution dilutes out the magnetic moment of the system which is not accounted for in the Doniach phase diagram.

\begin{figure}[th]
\centering
\includegraphics[width=1\linewidth]{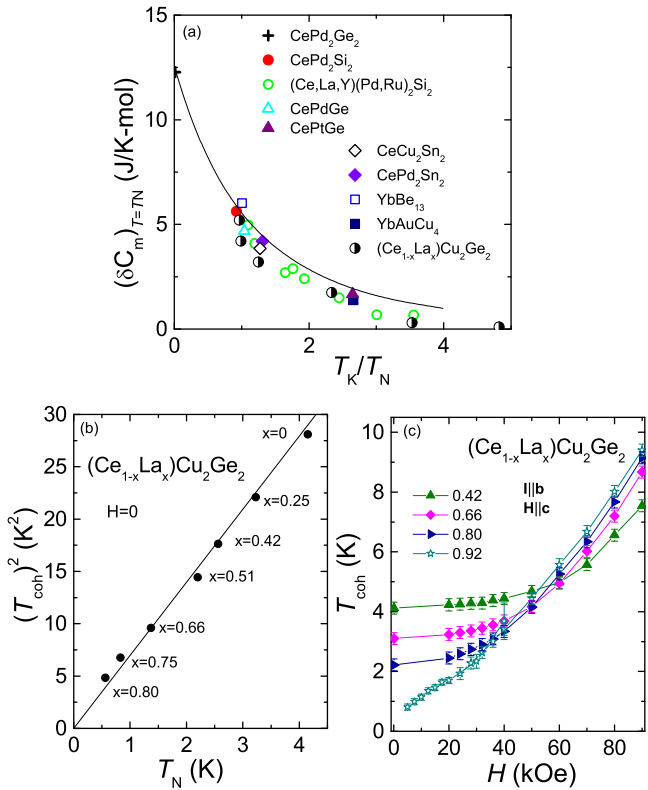}
\caption{\footnotesize (Color online) (a) Variation of the jump in the specific heat $\delta C_m$ at the magnetic transition as a function of $T_K/T_N$. The data for the compounds not studied in this work together with the solid curve were digitized from Ref. \cite{Besnus1992}. Solid line is the calculated specific heat jump at the $T_N$ for a doublet ground state system \cite {Besnus1992}. (b) $\left(T_{coh}\right)^2$ as a function of $T_N$. Solid line is a guide to the eye. (c) $T_{coh}$ as a function of $H$ (based on $\rho(T)$ at constant $H$ data given in the Supplemental Material \cite{SM}).}
\label{TT}
\end{figure}

Based on the molecular field calculations for the $S = 1/2$ resonant level model, a close relationship between the specific heat jump, $\delta C_m$, at the ordering temperature and the ratio between the two characteristic temperatures $T_K$ and $T_N$ for magnetic Ce and Yb Kondo systems with doublet ground states was found \cite{Besnus1992}. If the $T_K$-values shown in Fig. \ref{PDCe} are used and $T_K$ for CeCu$_2$Ge$_2$ assumed 4 K, (Ce$_{1-x}$La$_x$)Cu$_2$Ge$_2$ fits that description rather well, Fig. \ref{TT}(a). This further supports the thought that the estimated $T_K$ values are reasonable and the CEF ground state is a doublet.

The presence of the AFM transition and linear dependence of it on $x$ to a high value of La is not unique to the (Ce$_{1-x}$La$_x$)Cu$_2$Ge$_2$ system. Such behavior of $T_N$ upon La-dilution was also observed in (Ce$_{1-x}$La$_x$)Pd$_2$Si$_2$ \cite{Sampathkumaran1989}, as well as in (Ce$_{1-x}$La$_x$)Au$_2$Si$_2$, and (Ce$_{1-x}$La$_x$)Ag$_2$Si$_2$ \cite{Garde1994}. The Ce-based parent compounds of these families, including CeCu$_2$Ge$_2$, order antiferromagnetically, with different ordering wave vectors, and belong to the same $I$4/$mmm$ space group of the tetragonal crystal structure. However, a progression of the $T_{coh}$, that corresponds to the crossover from incoherent to coherent scattering, with La substitution was not commented on for these systems, perhaps because $T_N$ and $T_{coh}$ could not be well separated. In this respect, (Ce$_{1-x}$La$_x$)Cu$_2$Ge$_2$ appears to be a unique system $-$ the $T_{coh}$ is well separated from the AFM feature and extends all the way to $\sim$ 0.9 of La.

Remarkably, $\left(T_{coh}\right)^2$ is linearly proportional to $T_N$, Fig. \ref{TT}(b), over wide range of $x$ and both seem to go to zero at $x\sim$ 0.9. As of yet, there is no theory to explain a clear and compelling dependence of $T_{coh}$ on $T_N$.

In addition, a different field-dependence of the $T_{coh}$-value was found in the single-ion regime, Fig. \ref{TT}(c), which also supports the conclusion that $y\leq$ 0.09 defines the limit of the single ion regime in the zero-field limit (Fig. \ref{TT}(c) is based on the data given in the Supplemental Material \cite{SM}). The functional dependence of $T_{coh}$ on $H$ for $x$ = 0.92 is clearly different from that for smaller La concentrations, \textit{ i. e.}, for $x<$ 0.9, $T_{coh}$ saturates to a finite value as $H\to$ 0, and this is not the case for the $x$ = 0.92 data. Also, for $x$ = 0.92,  there is no $T_{coh}$ at $H$ = 0 and $T_{coh}$ is induced by magnetic field for all applied fields as opposed to smaller La concentrations.

This dilution study raises a number of questions/challenges for theories of the Kondo lattice. In particular, why do $T_N$ and $T_{coh}$ persist out to 90$\%$ La-substitution? And why does $T_N$ scale as $(T_{coh})^2$? In a simple percolation picture, this persistence indicates that the Kondo lattice has a low percolation threshold, consistent with a three-dimensional network with further neighbors; \textit{e.g.} - the cubic lattice with second and third neighbor interactions has a percolation threshold of 0.0976 \cite{Domb1966,Kurzawski2012}. Once coherence is established, the system can develop an AFM transition. At a more qualitative level, given that the clear AFM ordering signatures persist out to $x$ = 0.8, indicating that there is clear coupling and interaction between the remains of the Ce-sublattice, it is not at all surprising that this same coupling and interactions support coherence between ions. More sophisticated numerical studies of the dilute Kondo lattice give a crossover between coherent and single-ion Kondo behaviors at $x \approx 0.1$ only for very low conduction electron carrier densities, $n_c \ll 1$ \cite{Kaul2007,Burdin2013}. There are no experimental indications that $n_c$ for CeCu$_2$Ge$_2$ is so small, and this is contraindicated by the observation that $T_{coh} > T_K$ \cite{Burdin2000} and by bandstructure calculations \cite{Zwicknagl2007}. The large discrepancy between these numerical studies and our results suggests that large intersite correlations are essential to Kondo coherence in the dilute limit of (Ce$_{1-x}$La$_x$)Cu$_2$Ge$_2$, unlike in other materials such as (Ce$_{1-x}$La$_x$)CoIn$_5$ \cite{Nakatsuji2002,Nakatsuji2004} and (Ce$_{1-x}$La$_x$)Pb$_3$ \cite{Lin1987,Schlottmann1989}. The unusual scaling of $T_N$ with $(T_{coh})^2$ is unexpected and counters results of the two-fluid model, where $T^*$ and $T_N$ are expected to behave similarly \cite{Nakatsuji2004}.

In summary, La substitution drives $T_N$ in a roughly linear fashion from $\sim$ 4 K (for $x$ = 0 ) to below 0.36 K, the base temperature of our measurements, for $x>$ 0.8. However, $T_{coh}$, corresponding to the crossover from incoherent to coherent scattering, was observed up to $x\sim$ 0.9. This indicates that the percolation limit of the lattice of Ce ions is rather small and implies the 3D nature of the Kondo ``clouds". No non-Fermi liquid or Fermi liquid behavior that would indicate a quantum critical point (QCP) was observed in the thermodynamic and transport measurements upon suppression of $T_N$. We find $y\leq$ 0.09 is the single ion regime with $T_{coh}$ showing different behavior as a function of $H$ for $x>$ 0.9 and $x<$ 0.9. Remarkably, $(T_{coh})^2$ at $H$ = 0 was found to be linearly proportional to $T_N$ over wide range of $x$. (Ce$_{1-x}$La$_x$)Cu$_2$Ge$_2$ appears to be the only system where $T_{coh}$ is observed down to $x$ = 0.9 of La, $T_{coh}$ is well separated from magnetic ordering and single impurity effects, and $T_{coh}$ shows a parabolic dependence on $T_N$. Our results indicate that (Ce$_{1-x}$La$_x$)Cu$_2$Ge$_2$ is particularly compelling system and may be very useful for understanding the Kondo and RKKY effects.
 
\textit{Acknowledgment} The authors would like to thank J. Schmalian, P. Riseborough, Z. Fisk, B. C. Sales, J. D. Thompson, and F. Steglich for insightful discussions. This work was supported by the U.S. Department of Energy (DOE), Office of Science, Basic Energy Sciences, Materials Science and Engineering Division. The research was performed at the Ames Laboratory, which is operated for the U.S. DOE by Iowa State University under Contract $\#$ DE-AC02-07CH11358. E.D. M and H. K. were supported by the AFOSR-MURI grant No. FA9550-09-1-0603.

\end{document}

\title{Supplementary Material: Remarkably robust and correlated coherence and antiferromagnetism in (Ce$_{1-x}$La$_x$)Cu$_2$Ge$_2$}

\author{H. Hodovanets$^{1,2}$, S. L. Bud'ko$^{1,2}$, W. E. Straszheim$^{1}$, V. Taufour$^{1,2}$, E. D. Mun$^{1,2}$, H. Kim$^{2}$, R. Flint$^{1,2}$, and P. C. Canfield$^{1,2}$}

\affiliation{$^1$Ames Laboratory, Iowa State University, Ames, Iowa 50011, USA}
\affiliation{$^2$Department of Physics and Astronomy, Iowa State University, Ames, Iowa 50011, USA}

\maketitle
\subsection{Experimental}
Single crystals of (Ce$_{1-x}$La$_x$)Cu$_2$Ge$_2$ were grown from a ternary solution rich in Cu-Ge self-flux. The detailed description of the growth from the ternary melt can be found in Ref. \cite{Canfield1992,Canfield2010,Canfield2001}. A starting composition of (Ce$_{1-x}$La$_x$)$_{0.05}$Cu$_{0.475}$Ge$_{0.475}$ was placed in a 2-ml alumina crucible, sealed in a silica ampule under a small partial pressure of high purity argon gas. The ampule was heated to $1180\,^{\circ}\mathrm{C}$, dwelled there for 2 h, then cooled over 155 h to $825\,^{\circ}\mathrm{C}$, at which temperature the excess liquid was decanted using a centrifuge. To prepare the samples with nominal concentration of Ce, $y=1-x$, in the range 0.01$\leq y \leq$ 0.08, a master ingot of (Ce$_{0.08}$La$_{0.92}$)$_{0.05}$Cu$_{0.475}$Ge$_{0.475}$ was prepared first by arcmelting. Then, parts of the master ingot and La$_{0.05}$Cu$_{0.475}$Ge$_{0.475}$ were mixed in the ratios of 1:1, 1:2, and 1:4 to get the nominal values of $y$ = 0.04, 0.02, and 0.01, respectively. The single crystals grow as plates with the tetragonal $c$-axis perpendicular to the plate and the Laue back-reflection pattern (not shown here) confirmed that the plate-like samples have edges along (100) or (010) with the $c$-axis perpendicular to the plates. 

\begin{figure}[!b]
\centering
\includegraphics[width=1\linewidth]{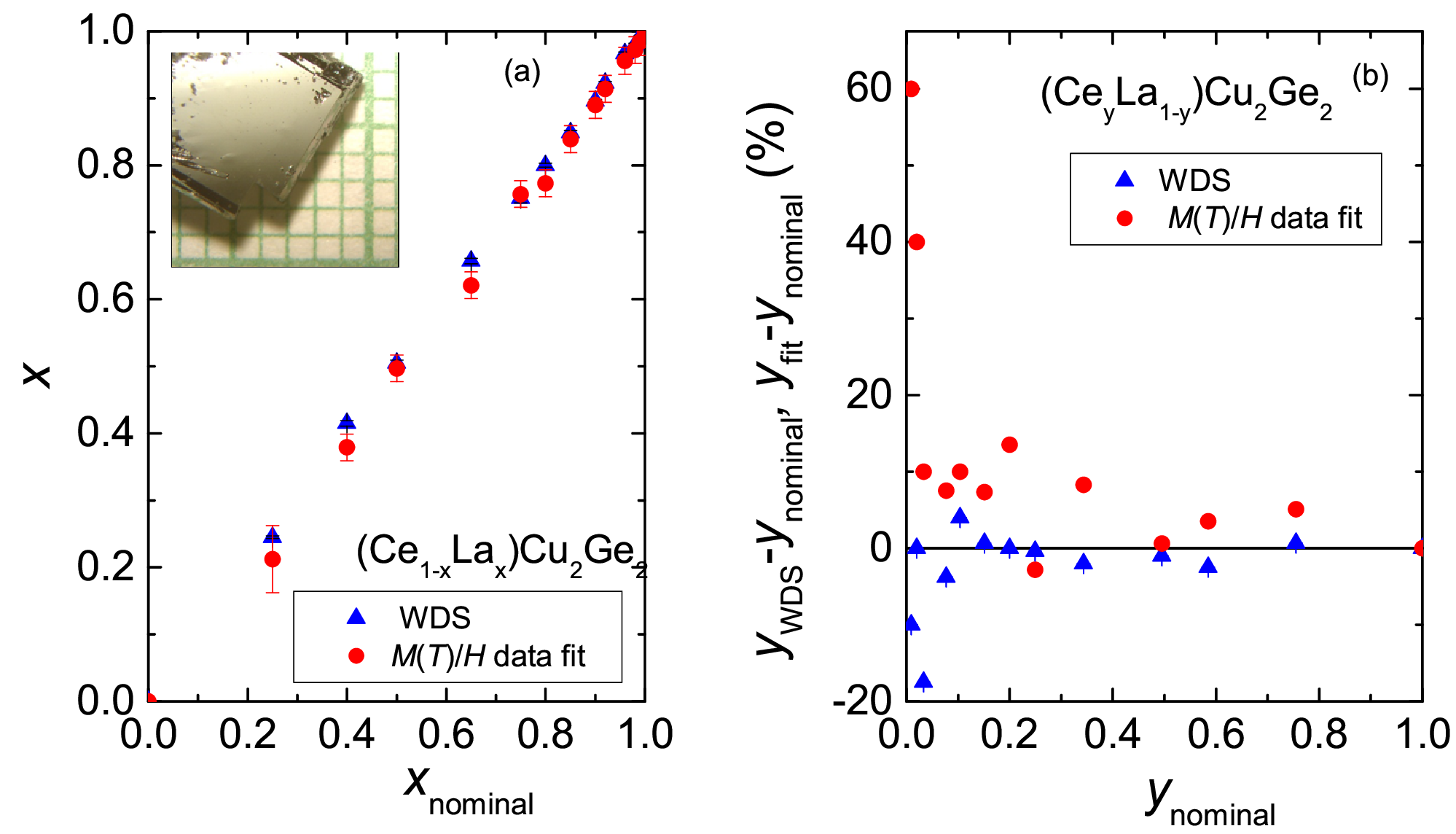}\renewcommand{\thefigure}{S\arabic{figure}}
\caption{\footnotesize (Color online) (a) La-concentrations of (Ce$_{1-x}$La$_x$)Cu$_2$Ge$_2$ single crystals, as determined by wavelength dispersive spectroscopy WDS and Curie-Weiss law fit of $M(T)/H$ data, versus nominal La-values. The inset shows large and shiny single crystal from the series. (b) Difference between the values obtained by WDS and Curie-Weiss law fit and nominal values as a function of Ce concentration, $y$.}

\label{WDS-FIT}

\end{figure}

The actual (as opposed to nominal) La concentration was determined in two ways (i) by wavelength dispersive x-ray spectroscopy (WDS) in a JEOL
JXA-8200 electron microprobe and (ii) based on the results of the Curie-Weiss fit. Figure \ref{WDS-FIT}(a) presents the results of both approaches. The modified Curie-Weiss law fit $\chi$ =$\chi _0$+$yC/(T-\theta$) of the polycrystalline average $\chi_{ave}=(2\chi_a+\chi_c)/3$ was performed in 150 K$<T<$ 300 K range. $y$ is Ce concentration and $y$ = 1$-x$. $C=(N_Ap_{eff}^2)/3k_B$, $p_{eff}=2.54\mu_B$ is the expected effective moment for the Ce$^{3+}$, $\mu_B$ is the Bohr magneton, $N_A$ is the
Avogadro number, and $k_B$ is the Boltzmann constant. Before the fit was performed, the $M(T)/H$ data of polycrystalline average of LaCu$_2$Ge$_2$ were subtracted. The agreement between these two ways of estimation of Ce concentrations is quite good, especially for larger amounts of Ce, Fig. \ref{WDS-FIT}(b). Clearly, uncertainty at the $\Delta y$ = 0.01$-$0.02 level becomes very large in terms of the $y$ value as $y$ becomes smaller than 0.2 or 0.1. On the other hand, the linearity of the $x-x_{nominal}$ plot is consistent with complete solubility of Ce in La for this structure. The WDS values of La concentrations will be used throughout the text if not specified otherwise. 

\begin{figure}[!t]
\centering
\includegraphics[width=1\linewidth]{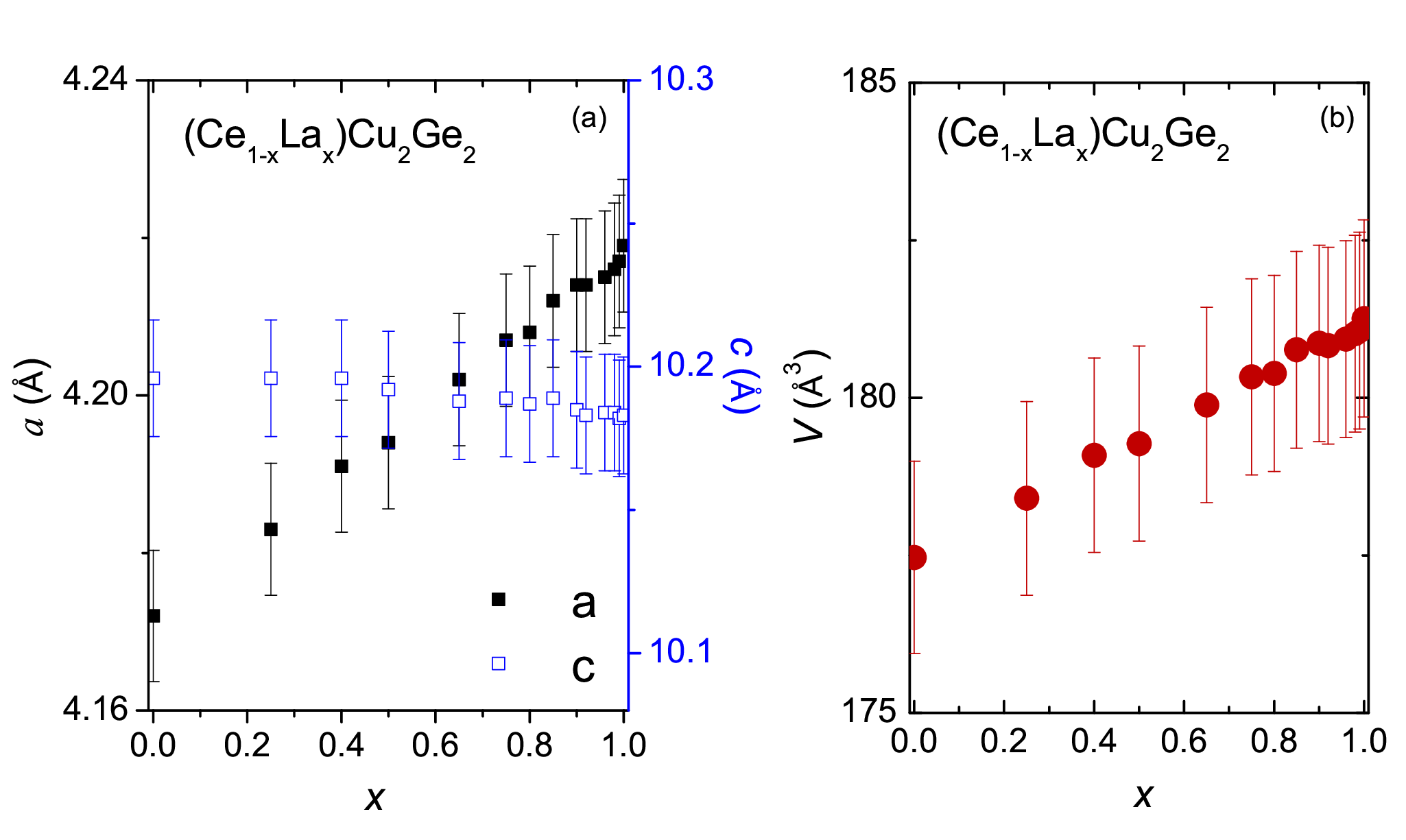}\renewcommand{\thefigure}{S\arabic{figure}}
\caption{\footnotesize (Color online) (a) Lattice parameters $a$ and $c$, and (b) the unit cell volume $V$ as a function of La concentration of (Ce$_{1-x}$La$_x$)Cu$_2$Ge$_2$ single crystals.}
\label{V}
\end{figure}

Powder x-ray diffraction (XRD) data were collected on a Rigaku MiniFlex diffractometer (using Cu $K_{\alpha 1,2}$ radiation) at room temperature. The x-ray measurements confirmed  $I4/mmm$ crystal structure of the samples. Lattice parameters were refined by the LeBail method using Rietica software. Figure \ref{V} shows the evolution of the lattice parameters $a$ and $c$ and a unit cell volume $V$ with increasing of La content. La-substitution results in the slight decrease of the value of $c$ (0.1$\%$) and increase of $a$ (1.1$\%$) lattice parameters. The unit cell volume $V$ increases with La substitution: $V$(LaCu$_2$Ge$_2$) is 2 $\%$ larger than $V$(CeCu$_2$Ge$_2$). 

Magnetic measurements were carried out in a Quantum Design Magnetic Property Measurement System (MPMS) SQUID magnetometer. For the measurements with $\bf{H}\| \bf{a}$, magnetic field along the plate, the samples were mounted in between two transparent plastic straws \cite{MPMSold}. For the magnetic field applied along the $\it{c}$-axis, perpendicular to the plates, the samples were mounted in the transparent plastic straws with two additional plastic straws holding the samples in place. This latter technique leads to the gap in the straw with a paramagnetic contribution to the total magnetic signal. The temperature dependence of the magnetization of the gap is almost temperature independent above $\sim$50 K, which contributes to the $\chi_0$ value, and linear magnetic field dependence. For the smallest Ce concentration, the fused silica rod with the gap in the middle was used for the magnetization measurements. The temperature and field dependent magnetization of the fused silica rod was measured beforehand to allow for the appropriate background subtraction.

Transport and specific heat measurements were performed in a Quantum Design Physical Property Measurement System PPMS-14 with $^3$He option and PPMS-9. A standard four-probe geometry, $\it ac$ technique ($\it f$=16 Hz, $\it I$=3$-$1 mA) was used to measure the electrical resistance of the samples. Electrical contact to the samples was made with platinum wires attached to the samples using EpoTek H20E silver epoxy. The care was taken to prepare the samples for the resistivity measurements so that the current was flowing along the edge of the plates, current along a (100)-direction. To calculate the resistivity of samples the distance between the midpoint of two voltage contacts and the cross-section area of the samples were used.  

For the specific heat measurements, a relaxation technique with fitting of the whole temperature response of the microcalorimeter was utilized. The background specific heat that includes sample platform and grease was measured for all necessary temperature values (in some instances magnetic field values as well) and subtracted from the total specific heat. The specific heat of LaCu$_2$Ge$_2$ was measured in the same temperature range and was used to estimate a phonon and ``non-correlated" electronic contributions to the specific heat of (Ce$_{1-x}$La$_x$)Cu$_2$Ge$_2$ single crystals.

The thermoelectric power (TEP) measurements were performed by a {\it dc}, alternating temperature gradient $\nabla T\|$\textbf{b} (two heaters and two thermometers) technique \cite{Mun2010b} with the temperature environment between 2 and 300 K provided by a Quantum Design PPMS. The samples were mounted with the help of Du-Pont 4929N silver paste directly on the surfaces of the SD packages of Cernox thermometers. The silver paste ensures thermal and electrical contact. The TEP measurements were extended down to $\sim$ 0.5 K by utilizing CRYO Industries of America, Inc. $^3$He cryostat. In this case, the samples were mounted on the surfaces of the SD packages of Cernox thermometers with the silver epoxy and were allowed to cure at $\sim$ $70\,^{\circ}\mathrm{C}$ for about one hour.

\subsection{Magnetization measurements}

\begin{figure}[!tbh]
\centering
\includegraphics[width=1\linewidth]{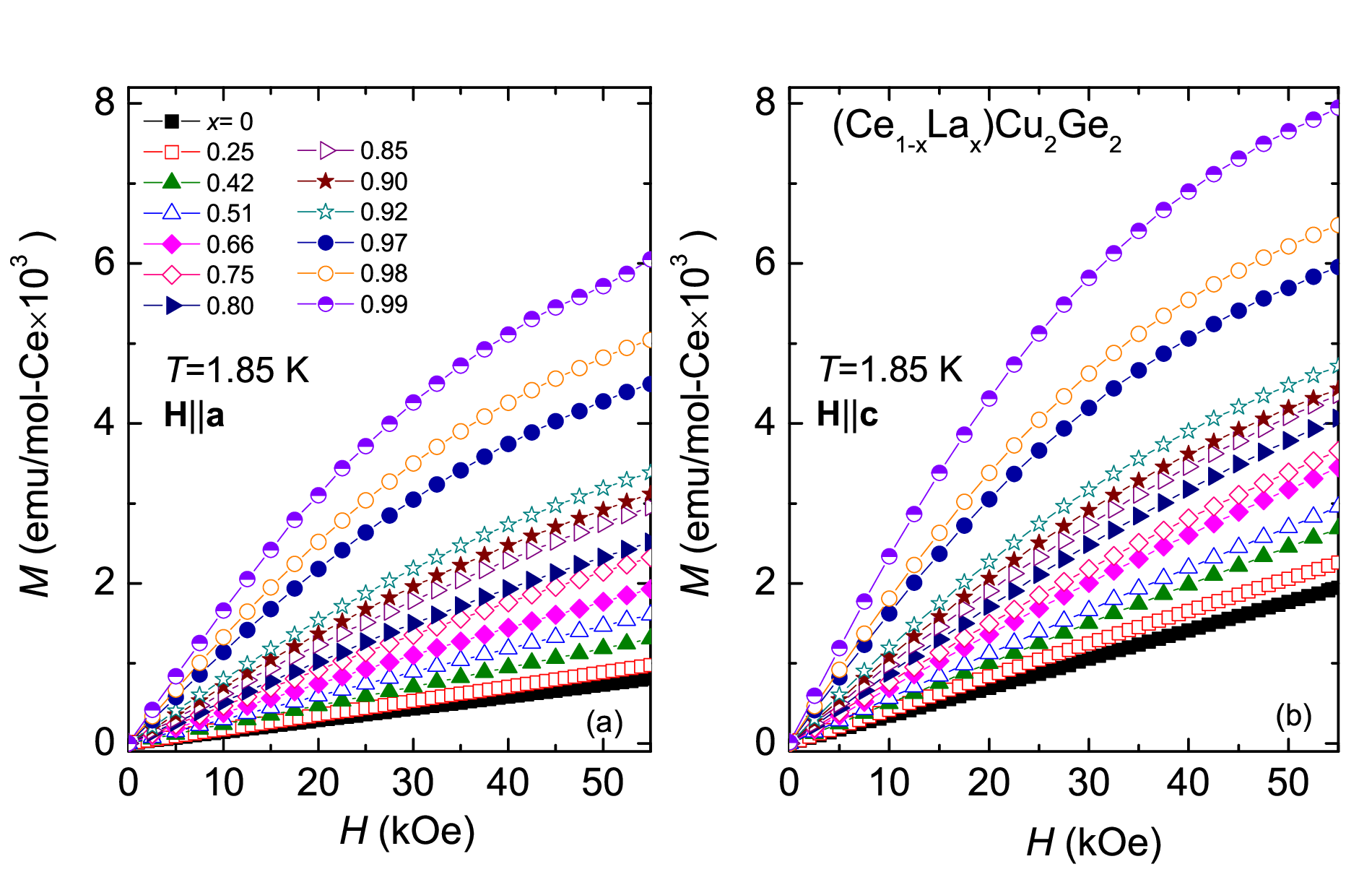}\renewcommand{\thefigure}{S\arabic{figure}}
\caption{\footnotesize (Color online) $M$ per mole of Ce versus $H$ at 1.85 K of (Ce$_{1-x}$La$_x$)Cu$_2$Ge$_2$ single crystals (a) \textbf{H}$\|$\textbf{a} and (b) \textbf{H}$\|$\textbf{c}.}
\label{MH-122}
\end{figure}

\begin{figure}[!tbh]
\centering
\includegraphics[width=1\linewidth]{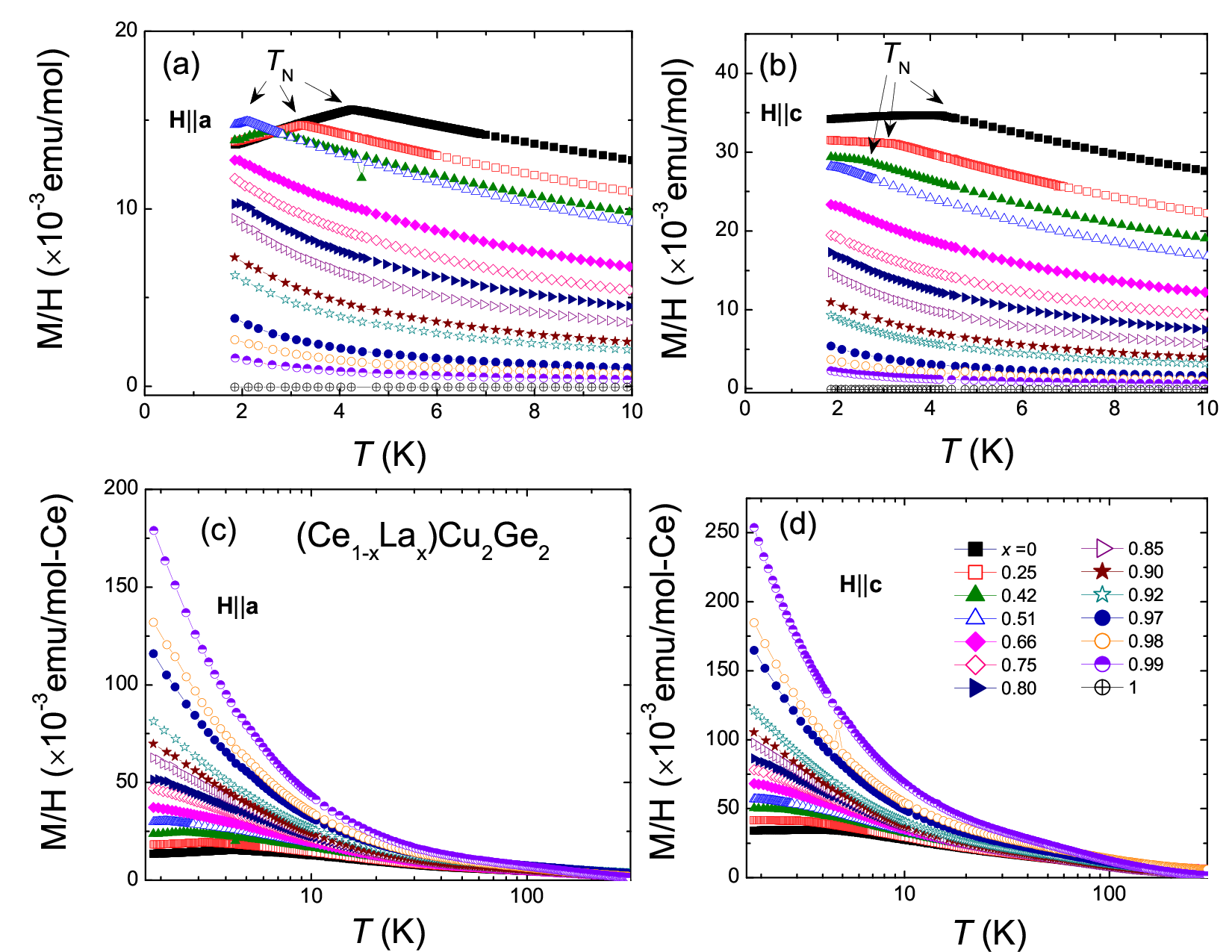}\renewcommand{\thefigure}{S\arabic{figure}}
\caption{\footnotesize (Color online) $M(T)/H$ of (Ce$_{1-x}$La$_x$)Cu$_2$Ge$_2$ single crystals (a) and (c) \textbf{H}$\|$\textbf{a}; (b) and (d) \textbf{H}$\|$\textbf{c}. For all samples containing Ce, a magnetic field of $H$ = 1 kOe was used to collect the data. For LaCu$_2$Ge$_2$, the field of $H$ = 10 kOe was used to collect the data. Panels (c) and (d) show the data, with $M(T)/H$ data of LaCu$_2$Ge$_2$ subtracted, in units of emu/mol-Ce, as opposed to emu/mol shown in panels (a) and (b). Hence, panels (c) and (d) do not have the data for $x$ = 1. Since the same, assigned to each concentration, symbols were followed in four panels, all legends are listed once in panel (d) for convenience.}
\label{MT-122}
\end{figure}

Figure~\ref{MH-122} shows magnetic isotherms $M(H)$ at $T$ = 1.85 K. $M(H)$ data were taken with the magnetic field applied along the $a$- and $c$- axes. The $M(H)$ data of LaCu$_2$Ge$_2$ were subtracted from each curve in an attempt to account for non-4$f$-shell related background. The magnetization for the field along the $c$-axis is larger than that for \textbf{H}$\|$\textbf{a} even for the smallest amount of Ce.

$M(H)$ data for 0 $\leq x\leq$ 0.51 manifest linear field dependence up to 55 kOe, except for $x$ = 0.51, $\textbf{H}\|\bf{c}$, where it deviates slightly from linearity. For higher La level samples, as the amount of La increases, the $M(H)$ data show a tendency to saturation at higher fields.
 
Figure \ref{MT-122} shows $M(T)/H$ for (Ce$_{1-x}$La$_x$)Cu$_2$Ge$_2$ single crystals for the magnetic field applied along the $a$- and $c$- axes. The AFM ordering for 0 $\leq x\leq$ 0.51 can be seen as a kink, much sharper for $\bf{H}\|\bf{a}$, that moves to lower temperature as the amount of La increases. The magnetization for $\bf{H}\|\bf{c}$ is larger than that for $\bf{H}\|\bf{a}$, consistent with $M(H)$ data shown in Fig. \ref{MH-122}, with the difference being most pronounced at lower temperatures.

\subsection{Specific heat measurements}
The low temperature linear fit of the $C/T$ vs $T^2$ of LaCu$_2$Ge$_2$ results in $\gamma$ = 4 mJ mol$^{-1}$K$^{-2}$ (3.6 mJ mol$^{-1}$K$^{-2}$ was estimated based on the electronic structure calculations \cite{Zwicknagl2007}) and $\Theta_D$ = 350 K. Due to the AFM transition at $\sim$ 4 K, a $\gamma$ value is hard to estimate precisely for CeCu$_2$Ge$_2$. The low temperature linear fit over 24 to 30 K temperature range of the $C/T$ vs $T^2$ of CeCu$_2$Ge$_2$ results in $\gamma\approx$ 90 mJ mol$^{-1}$K$^{-2}$ and the value of $C_p/T$ at 0.39 K is 245 mJ mol$^{-1}K^{-2}$.

\begin{figure}[t]
\centering
\includegraphics[width=0.8\linewidth]{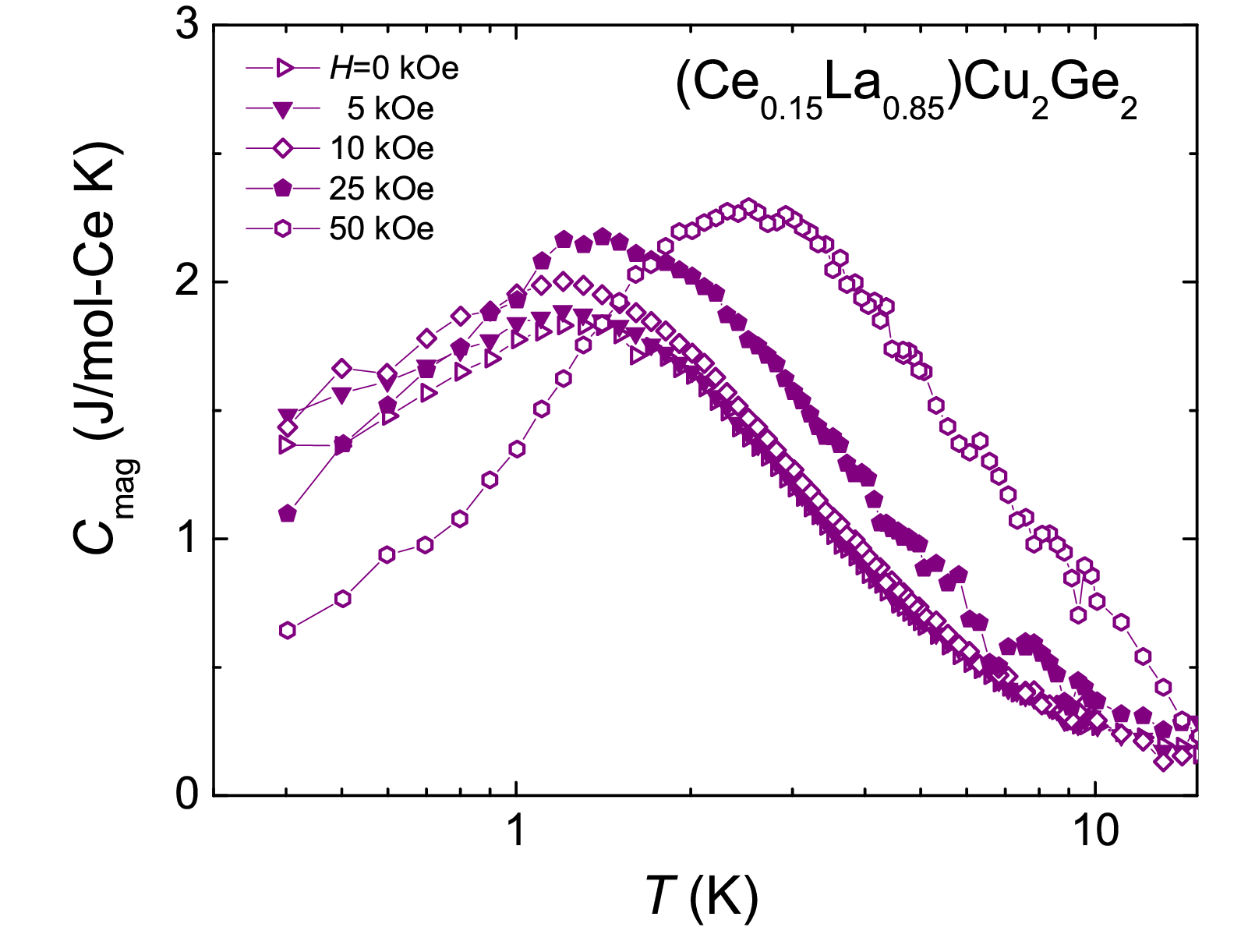}\renewcommand{\thefigure}{S\arabic{figure}}
\caption{\footnotesize (Color online) $C_{mag}(T)$ data of (Ce$_{0.15}$La$_{0.85}$)Cu$_2$Ge$_2$ single crystal for various applied magnetic fields.}
\label{CmH}
\end{figure}

To check the origin of the maximum in the specific heat, the $C_{mag}(T)$ data at various constant magnetic fields were collected for $x$ = 0.85, Fig. \ref{CmH}. These data can be compared with several possibilities:

(1) If the maximum in the specific heat is due to the first excited CEF level, then the application of the magnetic field results in the Zeeman splitting of the doublet which broadens the maximum.\cite{Hodovanets2013b} However, according to the specific heat \cite{Felten1987} and neutron scattering \cite{Loewenhaupt2012} measurements, the CEF splits the $J=\frac{5}{2}$ multiplet into three doublets with the first and second excited states being at $\Delta E_1\sim$ 197 K and $\Delta E_2\sim$ 212 K with respect to the ground state doublet. This rules out a Schottky anomaly in $C(T)$ data at low temperatures in zero field.

(2) In the case of spin glass state, the maximum in the specific heat is expected to broaden, decrease in height and move to higher temperatures upon application of magnetic field.\cite{mydosh1993} Although the feature does move up in temperature as the magnetic field is increased, it also increases in hight, this means that the spin-glass state is also unlikely.

(3) If the maximum in the specific heat is due to the single-ion Kondo effect, then upon application of magnetic field, as the magnetic field becomes comparable and larger than the energy $k_BT_K$, the specific heat peak becomes narrower and sharper and moves to higher temperatures.\cite{Schlottmann1985} Thus, the specific heat data of (Ce$_{0.15}$La$_{0.85}$)Cu$_2$Ge$_2$, Fig. \ref{CmH}, show similar behavior to that of single-ion Kondo impurity. The maximum seen in the specific heat for other concentrations of La is most likely associated with the $T_K$ of the single-ion Kondo impurity as well since La-substitution is not expected to drastically alter the CEF level scheme.  Similar behavior of the maximum in the specific heat data upon application of magnetic field was observed for heavily La diluted CeB$_6$ \cite{Gruhl1986}, CeCu$_6$ \cite{Satoh1989}, and CeAl$_2$ \cite{Rajan1982}. 

\begin{figure}[t]
\centering
\includegraphics[width=1\linewidth]{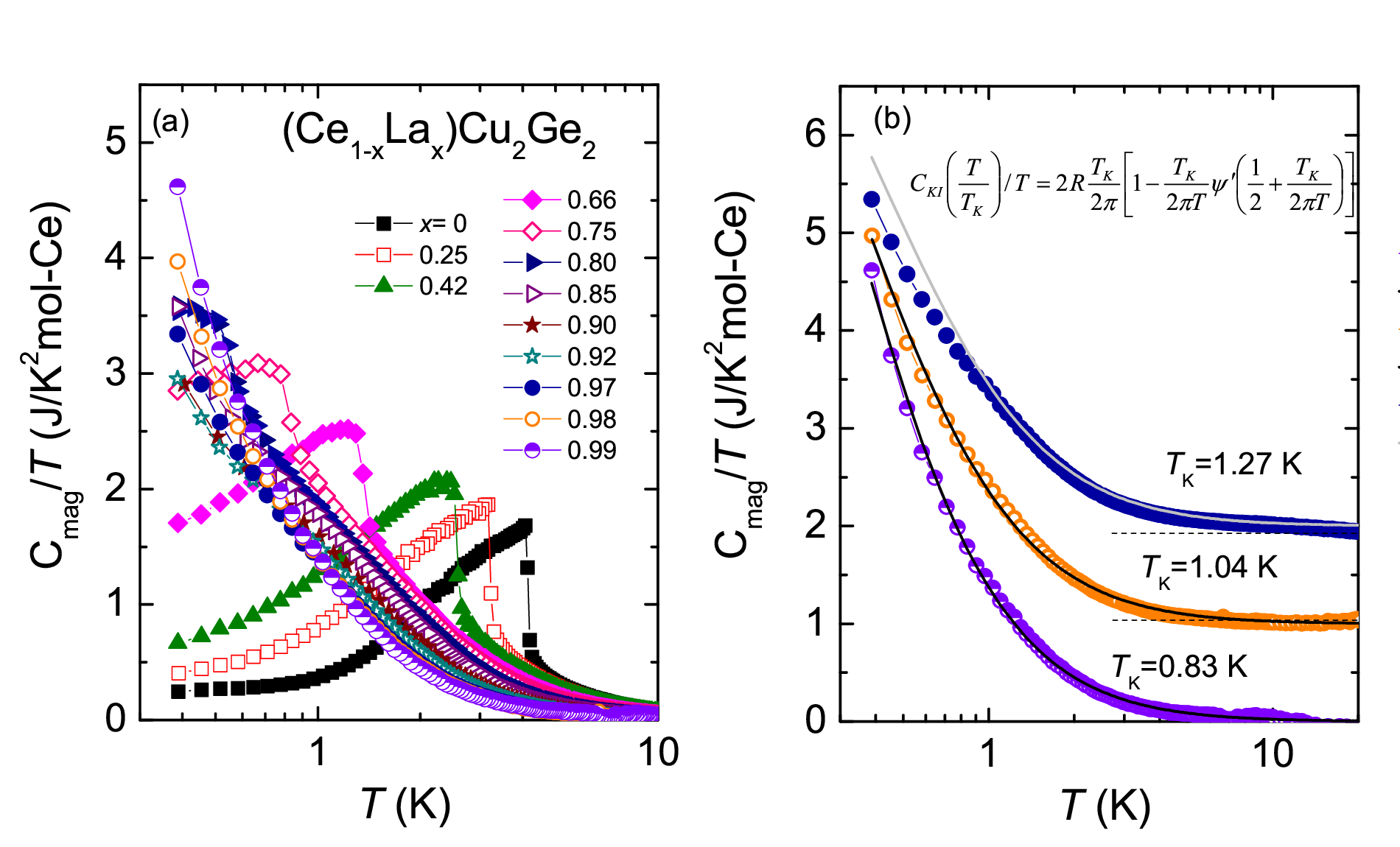}\renewcommand{\thefigure}{S\arabic{figure}}
\caption{\footnotesize (Color online) (a) $C_{mag}/T$ versus $T$ data normalized by the amount of Ce of (Ce$_{1-x}$La$_x$)Cu$_2$Ge$_2$ single crystals. (b) Fit of Kondo resonance model by Schotte and Schotte \cite{Schotte1975} for 1 mole of the Kondo impurity with the effective spin $S$ = $\frac{1}{2}$. $T_K$ is the Kondo temperature
defined as the width of the Lorentzian-shape Kondo resonance at the Fermi level. Each data set was offset from previous by 1 JK$^{-2}$(mol-Ce)$^{-1}$ for clarity.  Legends: $\bullet$ $x$ = 0.97, $\circ$ $x$ = 0.98, and half open circle $x$ = 0.99.}
\label{Cm-KI}
\end{figure} 

Figure \ref{Cm-KI}(a) shows $C_{mag}/T$ per mole of Ce atoms as a function of temperature for all samples measured. Again, a progression of the AFM ordering with the increase of La content is clearly seen for $x\leq$ 0.80. For 0.97 $\leq x\leq$ 0.99, the $\gamma\mid_{0.4 K}$ is rather large, indicating small Kondo temperature for these La concentrations. To check this assumption, the specific heat data were fitted based on the Kondo impurity model. The Kondo impurity with the effective spin $S$ = $\frac{1}{2}$ contribution to the specific heat per one mole of impurity according to the Kondo resonance model by Schotte and Schotte \cite{Schotte1975} is given by $$C_{KI}=2R\frac{T_K}{2\pi T}\left[1-\frac{T_K}{2\pi T}\psi' \left(1+\frac{T_K}{2\pi T}\right)\right],$$
where $R$ is the universal gas constant, $\psi'$ is the first derivative of the digamma function, and $T_K$ is the Kondo temperature
defined as the width of the Lorentzian-shape Kondo resonance at the Fermi level. The results of the fit to the $C_{mag}/T$ vs $T$ data for the three lowest Ce concentrations is given in Fig. \ref{Cm-KI} (b). $T_K$ is the only fitting parameter. The model seems to describe the data rather well for $x$ = 0.99 and 0.98. However, for $x$ = 0.97, this model does not describe the lowest temperature data well. Even worse fit (not shown here) is obtained for $x$ = 0.92. The presence of additional magnetic contribution to the specific heat can be one of the possible reasons as to why the model does not describe the data well for the larger amounts of Ce.

\subsection{Resistivity measurements}

\begin{figure}[tbh]
\centering
\includegraphics[width=1\linewidth]{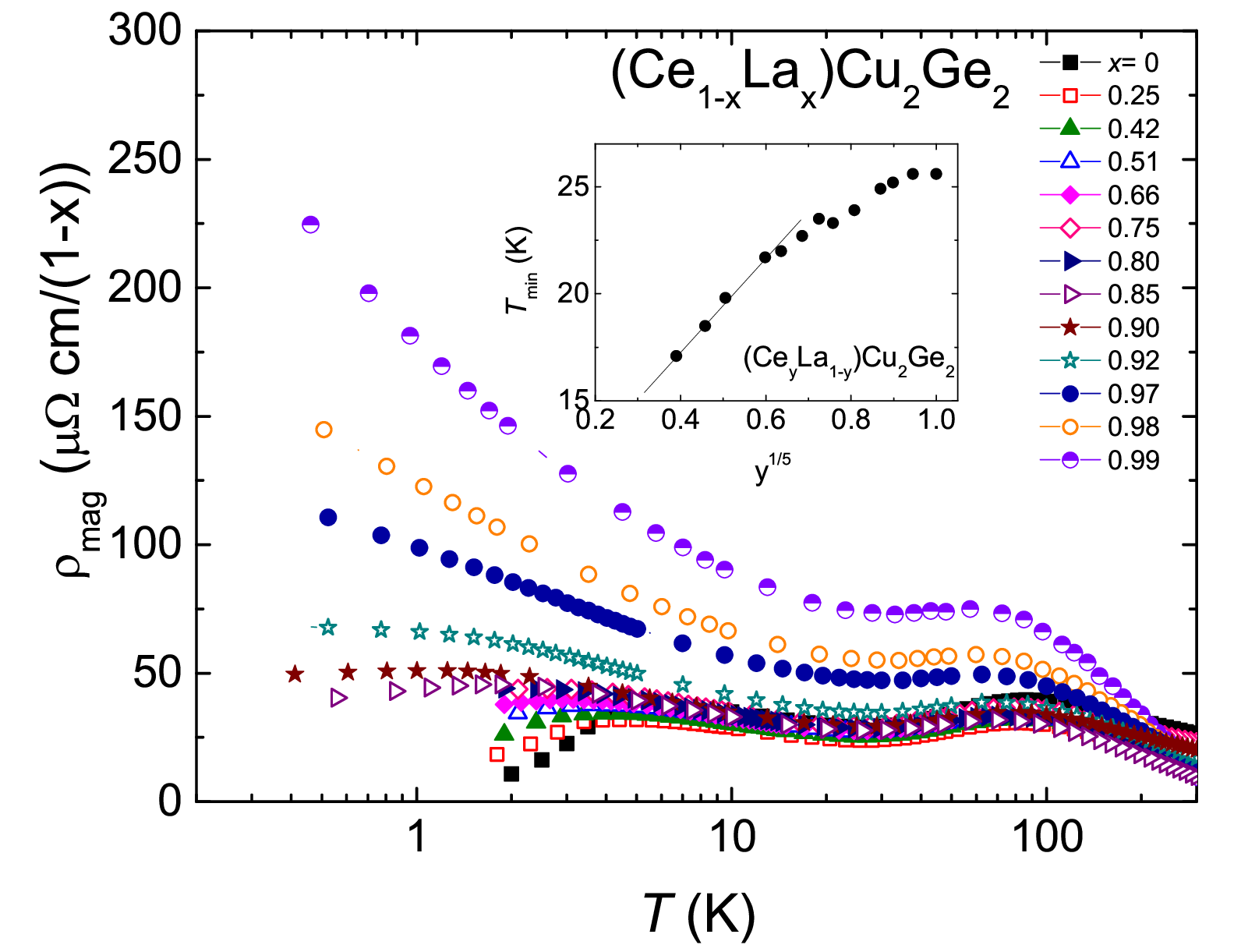}\renewcommand{\thefigure}{S\arabic{figure}}
\caption{\footnotesize (Color online) The zero-field, temperature-dependent resistivity (\textbf{I}$\|$\textbf{b}) $\rho(T)_{mag}$ data normalized to the Ce content of (Ce$_{1-x}$La$_x$)Cu$_2$Ge$_2$ single crystals on a semi-log plot. Every sixth data point is shown for clarity of the data presentation. The inset shows $T_{min}$ as a function of $y^{1/5}$.}
\label{Rmag}
\end{figure}

\begin{figure}[tbh]
\centering
\includegraphics[width=1\linewidth]{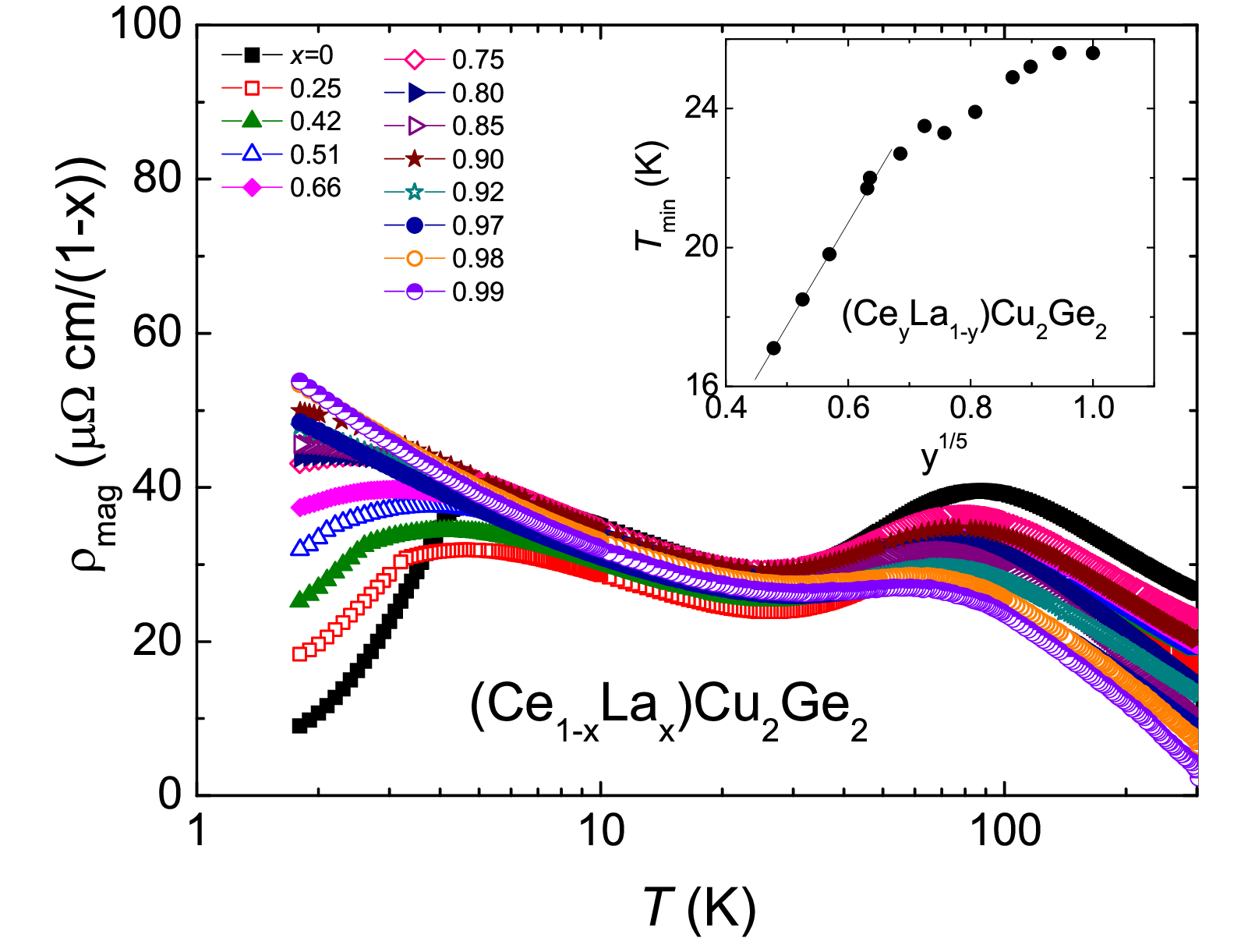}\renewcommand{\thefigure}{S\arabic{figure}}
\caption{\footnotesize (Color online) The zero-field, temperature-dependent resistivity (\textbf{I}$\|$\textbf{b}) $\rho(T)_{mag}$ data normalized to the Ce content of (Ce$_{1-x}$La$_x$)Cu$_2$Ge$_2$ single crystals on a semi-log plot. The four lowest Ce concentration were adjusted to $y$ = 0.025, 0.04, 0.06, and 0.10 instead of $y$ = 0.01, 0.02, 0.03, and 0.8, the data of which are shown in Fig. \ref{Rmag}, respectively. The inset shows $T_{min}$ as a function of $y^{1/5}$ with the four smallest $y$ values adjusted.}
\label{RmagCe}
\end{figure}

The magnetic contribution per Ce to the resistivity data $\rho_{mag}$ is shown in Fig. \ref{Rmag}. The data for $x\leq$ 0.90 seem to fall onto a single manifold for $T>T_N$. However, the data for $x\geq$ 0.92 show a clear departure from this trend. This behavior might be due to the greater variation of the $T_K$ for $x\geq$ 0.92 or the need to account for growing relative uncertainty in the Ce values. To account for this uncertainty, the $x$ values for the four lowest La concentration were adjusted so that all data collapse onto one manifold and the result is shown in Fig. \ref{RmagCe} together with the new graph of low-temperature $T_{min}$ versus adjusted values of $y^{1/5}$, here $y$ is concentration of Ce, that still holds for the new adjusted lowest concentrations of Ce. If these new values are adopted, then: (i) the $M(H)$ data at 55 kOe, shown in Fig. \ref{MH-122}, and $M(T)/H$ data at lowest temperatures, Fig. \ref{MT-122}, for these four lowest Ce concentrations fall in between the values of those for 0.42 $\leq x\leq$ 0.80; (ii) the fits of $C_{mag}/T$, shown in Fig. \ref{Cm-KI}, do not describe the data very well any longer and result in $T_K$ = 0.15, 0.21, and 0.21 K starting from the smallest Ce concentration. If additional parameter that reflects Ce concentration is included in the Schotte and Schotte model, then original values of Ce concentrations and $T_K$-values shown in Fig. \ref{Cm-KI} are recovered. Therefore, for now, the analysis of the data will be based on the values of Ce obtained by the WDS analysis.

\begin{figure*}[p]
\centering
\includegraphics[width=0.8\linewidth]{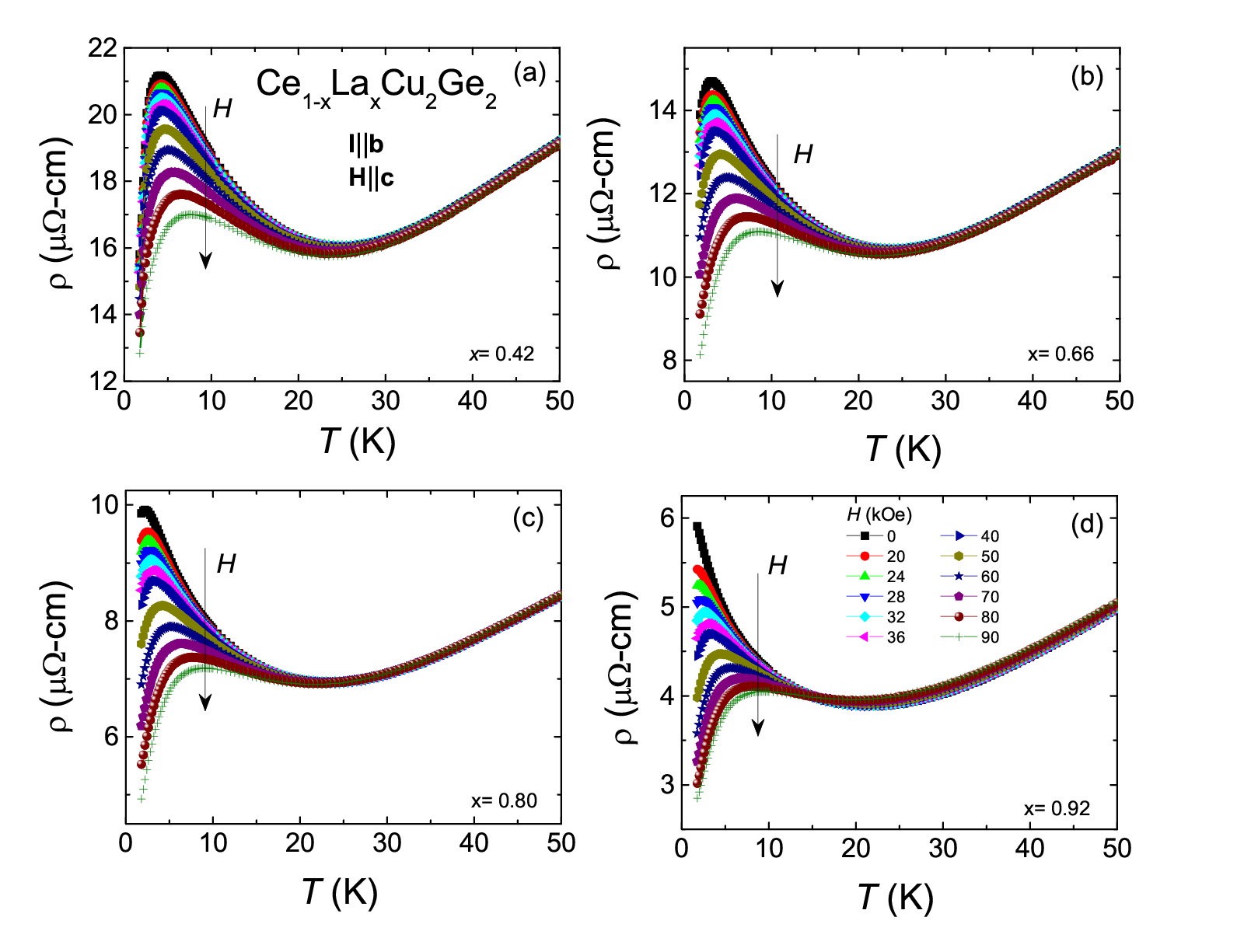}\renewcommand{\thefigure}{S\arabic{figure}}
\caption{\footnotesize (Color online) Temperature-dependent resistivity data at constant magnetic fields with \textbf{H}$\|$\textbf{c}, for (a) $x$ = 0.425, (b) $x$ = 0.66, (c) $x$ = 0.80, and (d) $x$ = 0.92 (\textbf{I}$\|$\textbf{b}).}
\label{cohHc}
\end{figure*}

\subsection{Resistivity in constant magnetic fields} 

The $\rho(T)$ data in constant magnetic fields, \textbf{H}$\|$\textbf{c}, for $x$ = 0.42, $x$ = 0.66, $x$ = 0.80, and $x$ = 0.92, are shown in Fig. \ref{cohHc}. The ln$T$ divergence of the resistivity at low temperatures is eliminated by the application of strong magnetic field. The strong magnetic field disallows spin-flip processes and causes a negative magnetoresistance.\cite{Ruvalds1988} The position of the $T_{max}$ is shifting toward higher temperatures as the magnetic field is increased. The data presented in Fig. \ref{cohHc} were used to create Fig. 5(c) which describes the evolution of $T_{max}$ or $T_{coh}$ as a function of magnetic field.

\subsection{Thermoelectric power: LaCu$_2$Ge$_2$}

\begin{figure*}[tbh]
\centering
\includegraphics[width=0.5\linewidth]{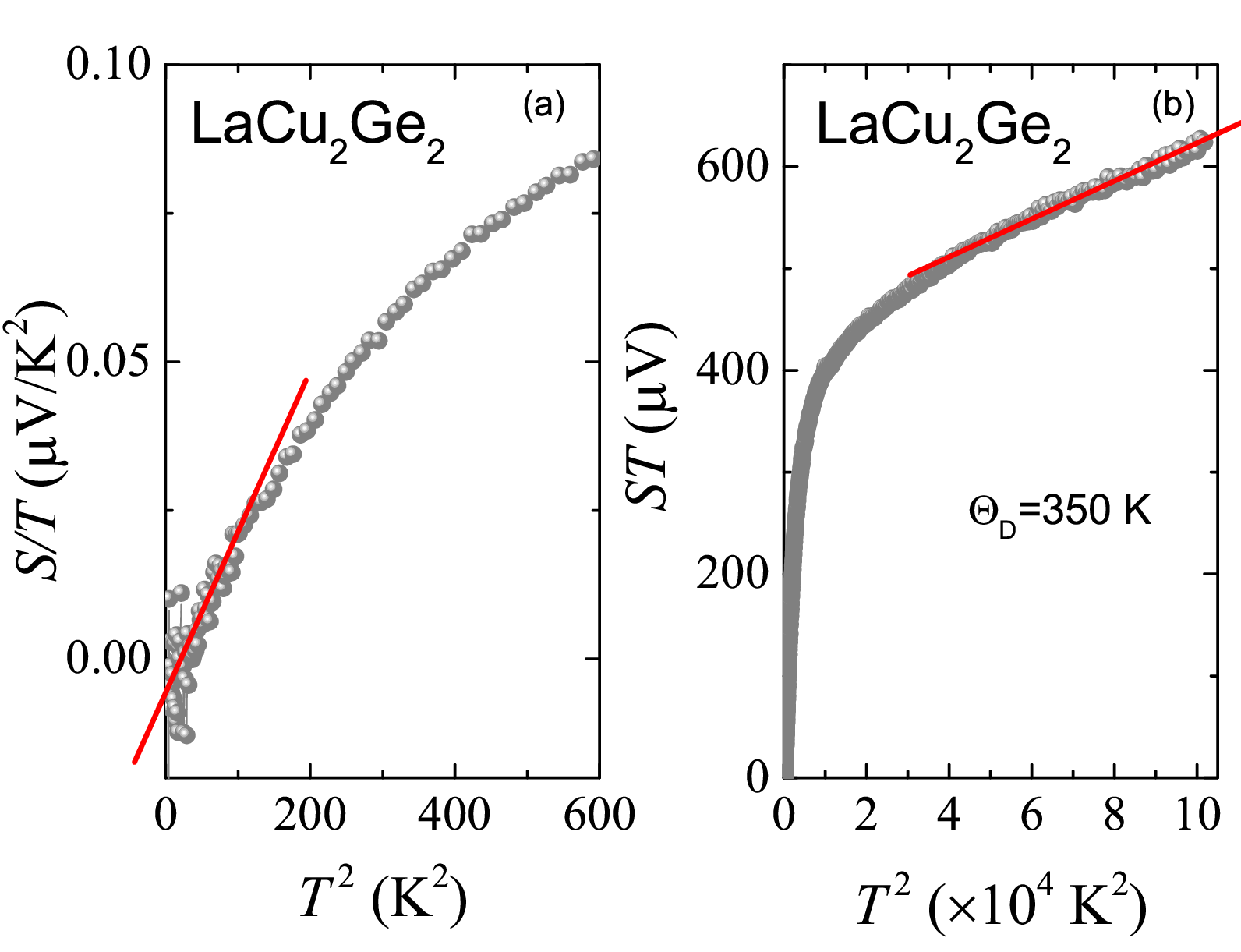}\renewcommand{\thefigure}{S\arabic{figure}}
\caption{\footnotesize (Color online) (a) $S/T$ vs $T^2$ and (b) $ST$ vs $T^2$ for LaCu$_2$Ge$_2$ single crystal. $\nabla T\|$\textbf{b}. Value of $\Theta_D$ is taken from the results of low-temperature fit of the $C/T$ vs $T^2$ data of LaCu$_2$Ge$_2$ (see section ``Specific heat measurements" above).}
\label{TEPLa}
\end{figure*}

A maximum in the thermoelectric power (TEP) data, due to the phonon drag contribution, is expected at 0.1$-$0.3$\Theta_D$ and for transition metals is in the range of 20-100 K.\cite{Blatt1976} Therefore the maximum observed at $\sim$ 70 K in the TEP data of LaCu$_2$Ge$_2$, Fig. 3, is probably due to the phonon drag contribution. In a simple model, as the temperature is increased from 0 K and $T\ll \Theta_D$, TEP can be written as $S=AT + BT^3$,\cite{Blatt1976} where the first term is electron contribution and the second term is phonon contribution. The $S/T$ plotted as a function of $T^2$ will result in a straight line. Fig. \ref{TEPLa} (a) shows $S/T$ vs $T^2$ measured on LaCu$_2$Ge$_2$ single crystal. The data deviate from linearity, this can be explained by the fact that the phonon relaxation processes have been ignored in the derivation of the above formula. 

At high temperatures, TEP maybe written as $S=A'T + B'/T$,\cite{Blatt1976} where the first term is electron contribution and the second term is phonon contribution. The $ST$ plotted as a function of $T^2$ will result in a straight line. Fig. \ref{TEPLa} (b) shows $ST$ vs $T^2$ measured on LaCu$_2$Ge$_2$ single crystal. Starting from $T\sim$ 260 K, the data do seem to behave linearly, however, measurement of the TEP up to much higher temperatures ($T>\Theta_D$) would be necessary to test this formula.

